\documentclass[preprint,showpacs,preprintnumbers,amsmath,amssymb]{revtex4}
\usepackage{graphicx}
\usepackage{dcolumn}
\usepackage{bm}
\usepackage[usenames]{color}
\newcommand{\be}{\begin{equation}}
\newcommand{\en}{\end{equation}}
\newcommand{\bea}{\begin{eqnarray}}
\newcommand{\ena}{\end{eqnarray}}

\begin{document}


\title{ Power law inflation with a non-minimally coupled scalar field in light of Planck and BICEP2 data: The exact versus slow roll results }

\author{Sergio del Campo\footnote{Deceased}}
\affiliation{ Instituto de F\'{\i}sica, Pontificia Universidad
Cat\'{o}lica de Valpara\'{\i}so, Avenida Brasil 2950, Casilla
4059, Valpara\'{\i}so, Chile.}

\author{Carlos Gonzalez}
\email{c.gonzalez@mail.ucv.cl} \affiliation{ Instituto de
F\'{\i}sica, Pontificia Universidad Cat\'{o}lica de
Valpara\'{\i}so, Avenida Brasil 2950, Casilla 4059,
Valpara\'{\i}so, Chile.}
\author{Ram\'on Herrera}

\email{ramon.herrera@ucv.cl} \affiliation{ Instituto de
F\'{\i}sica, Pontificia Universidad Cat\'{o}lica de
Valpara\'{\i}so, Avenida Brasil 2950, Casilla 4059,
Valpara\'{\i}so, Chile.}

\date{\today}
\begin{abstract}

We study power law inflation in the context of non-minimally coupled to the scalar curvature.
 We analyze the inflationary solutions  under an exact
 analysis  and also in the slow roll approximation. In both
 solutions, we  consider the recent data from Planck and BICEP2
 data to constraint the parameter in our model. We find that the slow
 roll approximation is disfavored  in the presence of  non-minimal
 couplings during the power law expansion of the Universe.
\end{abstract}

\maketitle

\section{Introduction}
It is well known that the inflationary scenario   has been an
important contribution to the modern cosmology, it was
particularly successful to explain cosmological puzzles such as
the horizon, flatness  etc. \cite{guth, linde1983}. As well, the
inflationary phase of the Universe provides an elegant mechanism
to elucidate  the large-scale structure\cite{est}, and also the
detected  anisotropy of the cosmic microwave background (CMB)
radiation\cite{CMB}.

On the other hand, the inflationary scenario  is supposed  to be
driven by a scalar field, and also this field can interact
fundamentally  with other fields, and in particular with the
gravity. In this form, is normal to incorporate an explicit
non-minimal coupling between the scalar field and the
gravitational sector. The non-minimal coupling with the scalar
Ricci, was in the beginning  considered in radiation problems in
Ref.\cite{rad}, and also in the renormalization of the quantum
fields in curved backgrounds, see Refs.\cite{r1,r2}. It is well
known, that scalar fields coupled with the
 curvature tensor arise  in different dimensions \cite{Capozziello:2011et}, and
 their importance  on cosmological scenarios  was studied for first time  in Ref.\cite{Jor}, together with
 Brans and Dicke \cite{BD}, although  also early the non-minimal coupling of the scalar field
 was analyzed in
 Ref.\cite{B1}.
In the context of the inflationary Universe, the non-minimal
coupling has been considered in Refs.\cite{val,fakir, faraoni},
and several inflationary models in the literature
\cite{Amendola:1999qq,Uzan:1999ch}. In particular, Fakir and Unruh
considered a new approach of the chaotic model from the
non-minimal coupling to the scalar curvature. Also,
 in Ref.\cite{futamase} considered the chaotic potential $V\approx\varphi^{n}$
$(n>4)$ for large $\varphi$ in the context to non-minimal
coupling, and found different constraints on the parameter  of
non-minimal coupling $\xi$ (see also  Ref.\cite{Bb2}). Recently,
the consistency relation for chaotic inflation model with a
non-minimal coupling to gravity was studied in Ref.\cite{new1},
and also a global stability analysis for cosmological models with
non-minimally coupled scalar fields was considered in
Ref.\cite{new2}.

On the other hand, in the context of the exact solutions, it
 can be
obtained for instance from a constant potential,  `` de Sitter''
inflationary model\cite{guth}. Similarly, an exact solution can be
found  in the case of  intermediate inflation model\cite{int},
however this inflationary model may be best considered from
slow-roll analysis. In the same way, an exact solution during
inflation can be achieved  from an exponential potential during
 ``power-law'' inflation in the case of General  Relativity. During
the  power law inflation, the scale factor is given by
$a(t)\propto t^p$, where the constant $p > 1$ \cite{pl}. In the
context of power law inflation with non-minimal coupling has been
studied in Refs.\cite{futamase,faraoni}. For power law inflation
with an effective potential $V\propto \varphi^n$ with $n>6$, was
analyzed in Ref.\cite{faraoni}. For this  inflationary model, it
can be found that only a very small range of the values of the
parameter $\xi$ is allowed for high values of the parameter $n$
(see also Refs.\cite{chino,Tsujikawa:2000tm}). Also, Futamase and
Maeda \cite{futamase} considered a chaotic inflationary scenario
in models having non-minimal coupling with curvature in the
context of power law inflation.

The main goal of the present work is to analyze the possible
actualization of an expanding power law inflation within the
framework of a non-minimal coupling with curvature, and how the
exact and slow roll solutions works in this theory. We shall
resort to the BICEP2 experiment data \cite{B2} and the Planck
satellite\cite{Ade:2013uln} to constrain the parameters in both
solutions. In particular, we obtain constraints on the fundamental
coefficients in our model.

The outline of the article is as follows. The next section
presents the basic equations and the exact and slow roll solutions
for our model. In Sect. III we determine the corresponding
cosmological perturbations. Finally, in Sect. IV we summarize our
finding. We chose units so that $c=\hbar=1$.

\section{Basic equations and exact versus slow roll solutions}

We start with the action for non-minimal coupling to gravity in
the Jordan frame  \cite{fakir2}

\begin{equation}
S=\int d^4x \sqrt{-g}\left[\left(\frac{1}{16\pi
G}+\frac{1}{2}\xi\varphi^{2}\right) R
+\frac{1}{2}g^{\mu\nu}\varphi_{;\mu}\varphi_{;\nu} - V
(\varphi)\right], \label{eq:1}
\end{equation}
where $G$ is the Newton`s gravitational constant, $\xi$ is a
dimensionless coupling constant, $R$ is the Ricci scalar and
$V$($\varphi$) is the effective  potential associated to the
scalar field $\varphi$. In particular, for the value  of the
coupling constant  $\xi=0$ corresponds to the minimal coupling,
and for the specific case in which $\xi=1/6$ is related to as
conformal coupling because the classical action possesses
conformal invariance. Also, different constraints on the parameter
$\xi$ can be found in the table of Ref.\cite{table}.

From the action given by Eq.(\ref{eq:1}), the dynamics in  a
spatially flat  Friedmann-Robertson- Walker (FRW) cosmological
model, is described by the equations

\begin{equation}
H^2=\frac{8\pi G}{3(1+8\pi
G\xi\varphi^2)}\left[\frac{1}{2}\dot{\varphi}^2+V-6\xi
H\varphi\dot{\varphi}\right],\label{H}
\end{equation}

\begin{equation}
\dot{H}=-\frac{8\pi G}{1+8\pi
G\xi\varphi^2}\left[\left(\frac{1}{2}+\xi\right)\dot{\varphi}^2-\xi(H\varphi\dot{\varphi}-\varphi\ddot{\varphi})\right],\label{dH}
\end{equation}
and
\begin{equation}
V_{,\,\varphi}= 6\xi\varphi (\dot{H}+2H^2)
-3H\dot{\varphi}-\ddot{\varphi},\label{df}
\end{equation}
where $H=\dot{a}/a$ is the Hubble parameter and  $a$  the scale
factor of the FRW metric. Dots means derivatives with respect to
time and  $V_{,\,\varphi}=\partial V(\varphi)/\partial \varphi$.

In order to obtain an exact solution, we will assume the power law
inflation, where the scale factor is characterized by $a \propto
t^p$, in which
 $p>1$. Here, the Hubble parameter $H=\dot{a}/a$ is given by $H(t)=p/t$.

Replacing  the scale factor $a \propto t^p$ in the
Eqs.(\ref{H})-(\ref{df}), we find an exact solution for the scalar
field, $\varphi$, given by

\begin{equation}
\varphi(t)=b\,t^{n}- \gamma,\label{sol2}
\end{equation}
where $b$, $n$  and $\gamma$ are constants. For an exact solution
of the scalar field, $\gamma$ is defined as
\begin{equation}
\gamma^2 =-\frac{1}{8\pi G \xi} =-
\frac{m_{p}^2}{\xi},\label{gamma}
\end{equation}
where $m_p$ is the reduced Planck mass and is defined as $m_p^2=(8\pi G)^{-1}$. Also,
in the following we will consider only
negative value of the parameter $\xi$ .

In order to obtain an exact solution, then the relations between
$p$ and $\xi$ with the exponent $n$ of Eq.(\ref{sol2}), are given
by

\begin{equation}
p=\frac{n^2-n}{2+n},\label{p}
 \end{equation}
and
\begin{equation}
\mid\xi\mid=\frac{(2+n)n}{2[n^2+3n-1]}.\label{e}
\end{equation}

Here, we note that the exponent $n$ of the solution of the scalar
field given by Eq.(\ref{sol2}), is such that, $n\neq -2$,
$n\neq(-3-\sqrt{13})/2\approx -3.3$ and
$n\neq(-3+\sqrt{13})/2\approx 0.3$. Considering that $p>1$, we
note from Eq.(\ref{p}), that  the value of the parameter $n$,
becomes $-2<n<1-\sqrt{3}\approx-0.73$ and $n>1+\sqrt{3}\approx
1.73$. In order to obtain the real roots of Eq. (\ref{e}),  we
considering only the value of $n<0 $. In this form, we find that
range for the parameter $\xi$ is given by
\begin{equation}
-\frac{1}{4+\sqrt{3}}<\xi< 0.\label{re1}
\end{equation}

The Hubble parameter as a function of the scalar field from
Eq.(\ref{sol2}),  becomes

\begin{equation}
H(\varphi)=p \,b^{\frac{1}{n}}(\varphi+\gamma)^{\frac{-1}{n}}.
\end{equation}

From Eqs.(\ref{H}) and (\ref{sol2}), the scalar potential as
function of the
 scalar field results
\begin{equation}
V(\varphi )=b^{\frac{2}{n}}(\varphi
+\gamma)^{1-\frac{2}{n}}(A\varphi - \gamma B),\label{pot}
\end{equation}
where the constants $A$ and $B$ are given by
\begin{equation}
A\equiv -3\mid\xi\mid p (p+2n)  - \frac{n^{2}}{2},
\end{equation}
and
\begin{equation}
B\equiv -3\mid\xi\mid p^{2} + \frac{n^{2}}{2}.
\end{equation}

From Eq.(\ref{pot}) we observe that the effective potential is
$V\neq 0$ for the value of $\varphi=0$, since $V(0) =
-Bb^{\frac{2}{n}}\gamma^{2-\frac{2}{n}}$. However, we note that
the constant $B$ is negative from the range of the parameter
$\xi$, see Eq.(\ref{re1}) , and then  the effective potential
becomes $V(\varphi=0)>0$ (see Refs.\cite{hosotani,faraoni} for
other
 $V(0)\neq 0$).

In order to reproduce the present value of the Newton's
gravitational constant, we can write from Ref.\cite{futamase},
that $G_{eff} = \frac{G}{1- \frac{\varphi^{2}}{\gamma^{2}}} $.
Here,  $G_{eff}$ corresponds to the effective Newton's
gravitational constant. By considering that $G_{eff}>0$, we
observe  that the scalar field is well supported by the condition
$\mid\varphi\mid < \frac{m_{p}}{\sqrt{\mid\xi\mid}}=\gamma$, then
the inflationary scenario  can be realized in the region in which
$-\gamma\lesssim \varphi\lesssim \gamma$.


In the following, we will study the power law solution in the slow
roll conditions.  Following Ref. \cite{Torres:1996fr}  the slow
roll approximation are defined as $\dot{H} \ll H^2$ and
$\ddot{\varphi}\ll 3H \dot{\varphi}$.
 In this form, the slow roll field equations from
Eqs.(\ref{H})-(\ref{df}) can be written as

\begin{equation}
H^2\simeq \frac{8\pi G}{3(1+8\pi G\xi\varphi^2)} ( V(\varphi)-6\xi
H\varphi\dot{\varphi} ),
\end{equation}
and
\begin{equation}
3H\dot{\varphi}+V_{,\,\varphi}\simeq 12\xi\varphi\, H^2 .
\end{equation}

Considering the power law expansion $a\propto t^p$, we get

\begin{equation}
\varphi(t) = bt^{-2} -\gamma,\label{solp2}
\end{equation}
with
$$
p=\frac{2}{\mid\xi\mid}-4>1,
$$
where  the constant $\gamma$ as before is given by
Eq.(\ref{gamma}), and during the slow roll approximation the value
of  $|\xi|<2/5$. As before, since the Hubble parameter is
$H\propto t^{-1}$, we can eliminate $t$ by using Eq.(\ref{solp2}),
thus $ H(\varphi)= pb^{-1/2} (\varphi +\gamma)^{1/2}, $ and the
effective potential results

\begin{equation}
V(\varphi)=  3pb^{-1}\mid\xi\mid(\varphi +\gamma)^{2}[ \gamma p +
\varphi(4-p)].\label{17}
\end{equation}

Here, we note that in the slow roll approximation, the solution
for the scalar field is $\varphi\propto t^{-2}$ (with the fixed
value $n=-2$, see Eq.(\ref{sol2})), and also the exact effective
potential reduces to the cubical polynomial form as potential
given by  Eq.(\ref{17}).

\section{Cosmological  perturbations}

In this section we will analyze the scalar and tensor
perturbations for our model. The general perturbation metric about
the flat background is given by
\begin{equation}
ds^{2}=-(1+2\Phi)dt^2+2a(t)\Theta_{,i}dx^idt+a^2(t)[(1-2\psi)\delta_{ij}+2E_{,i,j}+2h_{ij}]dx^idx^j,\label{me}
\end{equation}
where $\Phi$, $\Theta$, $\psi$ and $E$ represent to the
scalar-type metric perturbations, and the tensor $h_{ij}$
corresponds the transverse traceless perturbation.

On the other hand, the perturbation in the scalar field $\varphi$
is specified as $\varphi(t,\vec{x}
)=\varphi(t)+\delta\varphi(t,\vec{x} )$, where $\varphi(t)$ is the
background scalar field that satisfies the Eq.(\ref{df}),  and
$\delta\varphi(t,\vec{x} )$  is  a small perturbation that
represents small fluctuations of the corresponding scalar field.
In this form, we introduce comoving curvature perturbations, given
by \cite{Bardeen}

\begin{equation}
{\cal{R}}= \Psi +
\mathcal{H}\frac{\delta\varphi}{\varphi^{\prime}}\;,\label{RRRR}
\end{equation}
where now the Hubble parameter is defined as
$\mathcal{H}\equiv\frac{a^{\prime}}{a}$ and a prime denotes a
derivative with respect to a conformal time $d\eta=a(t)^{-1} dt$.

From  the action given by Eq.(\ref{eq:1}), we find that the
perturbed equations of motion are given by

\begin{equation}
\Psi^{\prime} + \mathcal{H} \Phi = \frac{1}{2F}(f^{\prime} - \Phi
F^{\prime}- \mathcal{H}f+\varphi^{\prime}\delta\varphi
),\label{p1}
\end{equation}
\begin{equation}
\Psi -\Phi =\frac{f}{F},\label{p2}
\end{equation}
and
\begin{equation}
3\mathcal{H}^{2}f+2F (\nabla^{2}\Psi -3\mathcal{H}[\Psi^{\prime} +
\mathcal{H}\Phi ]) = -3\mathcal{H} f^{\prime} +
3F^{\prime}\Psi^{\prime} + \nabla^{2}f +
\varphi^{\prime}\delta\varphi^{\prime}\label{p3}
\end{equation}
$$-\Phi\varphi^{\prime 2 }+ 6\Phi\mathcal{H}F^{\prime} + a^{2}
V_{,\varphi}\delta\varphi,
$$
where $F=F(\varphi)=(8\pi G)^{-1}+\xi\varphi^2$ and $f=\delta
F=(\partial F/\partial\varphi)\delta\varphi$. Also,  we
considering the longitudinal gauge in the perturbed metric
(\ref{me}), where $\Phi=\Phi(t,\vec{x})$ and
$\Psi=\Psi(t,\vec{x})$  are gauge-invariant variables introduced
in Ref.\cite{Bardeen}.

Defining two auxiliary functions; $\alpha=\frac{3F^{\prime
2}}{2F}+\varphi^{\prime 2}$ and
$\beta=\mathcal{H}+\frac{F^{\prime}}{2F}$, then using
Eq.(\ref{RRRR}), the equations (\ref{p1}),  (\ref{p2}),
 and (\ref{p3})  can be written in the form
\begin{equation}
{\cal{R}}^{\prime}+ A_1\delta\varphi + B_1\delta\varphi^{\prime} +
\beta {\cal{R}} =0,\label{ac1}
 \end{equation}
and
 \begin{equation}
C\delta\varphi^{\prime}-6F\beta {\cal{R}}^{\prime} +
\frac{CA_1}{B_1}\delta\varphi
 - C \varphi^{\prime} {\cal{R}} +
2F\nabla^{2}{\cal{R}} + 2FB_1\nabla^{2}\delta\varphi=0,\label{p4}
\end{equation}
 where $A_1$, $B_1$ and $C$ are given by
$A_1=\frac{\beta\varphi^{\prime\prime}}{\varphi^{\prime 2}}-
\frac{2\beta^{2}}{\varphi^{\prime}}$ ,
$B_1=-\frac{\beta}{\varphi^{\prime}}$ and
$C=\frac{1}{\varphi^{\prime}}(6F\beta^{2}-\alpha)$, respectively.

Following Ref.\cite{noh}, the Eqs.(\ref{ac1}) and (\ref{p4}) can
be decoupled, and then the equation of motion for the curvature
perturbation becomes

\begin{equation}
\frac{1}{a^{3}Q_{s}}\frac{d}{dt}{(a^{3}Q_{s}\dot{{\cal{R}}})} +
\frac{k^{2}}{a^{2}}{\cal{R}}=0,\,\,\,\,\,\,\mbox{with}\,\,\,\,\,Q_{s}\equiv\frac{\alpha}{\beta^{2}},\label{ff}
\end{equation}
 where $k$ is a comoving wavenumber. Here, we note that the equation for the curvature perturbation given by
 Eq.(\ref{p4}),  coincides with the equation obtained in
 Ref.\cite{hwang}. Introducing new variables, in which
 $z=a\sqrt{Q_s}$ and $v=a{\cal{R}}$, the above equation can be
 written as $v^{\prime\prime} + (k^{2}-\frac{z^{\prime\prime}}{z})v=
 0$, see Ref.\cite{shinji}.

As argued in Refs.\cite{noh,shinji}, the solution of the above
equation can be expressed by the combination of the Hankel
function, and  the scalar density perturbation ${\cal{P}}_S$,
could be written as

\begin{equation}
{\cal{P}}_{S}\equiv \frac{k^{3}}{2\pi^{2}}|{\cal{R}}|^{2} =
A_S^{2}\left[\frac{k |\eta |}{2}\right]^{3-2\nu_{s}},\label{pert}
\end{equation}
where $A_{S}^{2}\equiv \frac{1}{Q_{s}}(\frac{H}{2\pi})^{2}
(\frac{1}{aH|\eta|})^{2} [\frac{\Gamma(\nu_{s})}{\Gamma(3/2)}]^{2}
$ and $\nu_{s} \equiv\sqrt{\gamma_{s} + 1/4}$. Here, $\gamma_s$ is
defined as
$$
\gamma_{s} =\frac{(1+\delta_{s})(2-\epsilon +
\delta_{s})}{(1-\epsilon)^{2}},\,\,\,\,\,\mbox{where}\,\,\,
\;\delta_{s}\equiv\frac{\dot{Q_{s}}}{2HQ_{s}},\,\,\,\mbox{and}\,\,\,\epsilon\equiv-\frac{\dot{H}}{H^{2}}\,.
$$

The scalar spectral index $n_S$ is given by $n_S-1=\frac{d\ln
{\cal{P_{S}}}}{d\ln k}$. From Eq.(\ref{pert}), it follows that
\cite{shinji}

\begin{equation}
n_{S}=4-\sqrt{4\gamma_{s} +1}.
\end{equation}

The spectrum of tensor perturbations, $h_{ij}$ can be obtained  in
a similar way, since $h_{ij}$ satisfies the equivalent form of
Eq.(\ref{ff}). In this form, following Ref.\cite{shinji}
 in the which  $Q_{s} \to Q_{T}= F$,  the power spectrum of the
 tensor modes ${\cal{P}}_{T}$, can be written as

\begin{equation}
{\cal{P}}_{T}\equiv A_{T}^{2}\left[\frac{k |\eta
|}{2}\right]^{3-2\nu_{T}},\label{pt}
\end{equation}

where $A_{T}^{2}\equiv \frac{8}{Q_{T}}(\frac{H}{2\pi})^{2}
(\frac{1}{aH|\eta|})^{2}
[\frac{\Gamma(\nu_{T})}{\Gamma(3/2)}]^{2}$, and the parameters
$\nu_{T}$,  $\gamma_{T}$ and $\delta_{T}$   are given by
$$
\nu_{T} \equiv\sqrt{\gamma_{T} + 1/4},\,\,\,\gamma_{T}
=\frac{(1+\delta_{T})(2-\epsilon +
\delta_{T})}{(1-\epsilon)^{2}},\,\,\,\mbox{and}\,\,\,\delta_{T} =
\frac{\dot{Q}_{T}}{2HQ_{T}}.
$$
 Here, the index of the tensor perturbation is given by
${n}_{T}=3-\sqrt{4{\gamma}_{T} +1}$.

On the other hand, an essential  observational quantity is the
tensor to scalar ratio $r$, which is defined as
$r={\cal{P}}_T/{\cal{P}}_S$.   In this way, combining
Eqs.(\ref{pert}) and (\ref{pt}) the tensor to scalar ratio is
given by

\begin {equation}
r=
8\frac{Q_{s}}{Q_{T}}\left[\frac{\Gamma(\nu_{T})}{\Gamma(\nu_{s})}\right]^{2}=
8\left[\frac{\frac{3F^{\prime 2}}{2F}+\varphi^{\prime
2}}{F(\mathcal{H}+\frac{F^{\prime}}{2F})^{2}}\right]\left[\frac{\Gamma(\nu_{T})}{\Gamma(\nu_{s})}\right]^{2}.\label{r}
\end{equation}

In this form, considering the power law inflation, i.e., $a\propto
t^p$, and the exact solution for the scalar field given by
Eq.(\ref{sol2}), then the scalar spectral index $n_s$ as a
function of the scalar field $\varphi$  can be written as
\begin{equation}
n_S(\varphi)-1= 3-\sqrt{4\gamma_s(\varphi)+1},\label{nse}
\end{equation}
where  $\gamma_s(\varphi)$ is given by
$$
\gamma_s(\varphi)=p\,\left[\frac{(1+\delta_s(\varphi))\,(2p-1+p\delta_s(\varphi))}{(p-1)^2}\right],
$$
with
$$
\delta_{s}(\varphi)=\frac{n}{p}\left[1+(\gamma - \varphi)^{-1}
\left(\frac{6m_{p}^{2}\gamma^{2}\varphi}{\varphi^{2}(6m_{p}^{2}-\gamma^{2})+\gamma^{4}}+\frac{n\gamma(\varphi+\gamma)}{p\gamma-\varphi(p+n)}\right)\right].
$$

Also,  the tensor to scalar ratio $r$ can be written in terms of
the scalar field $\varphi$ as
\begin{equation}
r(\varphi)=\frac{8n^{2}}{m_{p}^{2}}\left(\frac{\varphi^{2}(6m_{p}^{2}-\gamma^{2})+\gamma^{4}}
{(p\gamma-\varphi(p+n))^{2}}\left[\frac{\Gamma(\nu_{T})}{\Gamma(\nu_{s})}\right]^{2}\right).\label{re}
\end{equation}
Here, we have considered Eqs.(\ref{sol2}) and (\ref{r}).

In the slow-roll approximation, following Ref.\cite{shinji}, the
scalar spectral index is given by
\begin{equation}
n_S-1\approx-2(\delta_s+\epsilon),\label{nsr}
\end{equation}
and the tensor to scalar ratio $r$ becomes
\begin{equation}
r\approx 16\left(\frac{\dot{F}}{2HF}+\epsilon \right),\label{rr}
\end{equation}
because during the slow roll approximation, $\mid\delta_s\mid< 1$
and $\mid\delta_T\mid< 1$. Also, here we considered that
$\Gamma(\nu_s)\simeq\Gamma(\nu_T)\simeq\Gamma(3/2)$, since that
both perturbations are  close to scale invariant \cite{shinji},
see also Refs.\cite{noh,kaiser}.

Under the slow roll approximation the quantity $\delta_s$ is given
by
$$
\delta_s=\frac{\mid\xi\mid}{2|\xi|-1} \left[1+(\gamma -
\varphi)^{-1}
\left(\frac{6m_{p}^{2}\gamma^{2}\varphi}{\varphi^{2}(6m_{p}^{2}-\gamma^{2})+
\gamma^{4}}-\frac{\mid\xi\mid\gamma(\varphi+\gamma)}{(1-2|\xi|)[\gamma-\varphi]+|\xi|\varphi}\right)\right].
$$
Here, we have considered  the slow-roll solution  for the scalar
field given by  Eq.(\ref{solp2}).

\begin{figure}[th]
{\hspace{-5
cm}\includegraphics[width=6.0in,angle=0,clip=true]{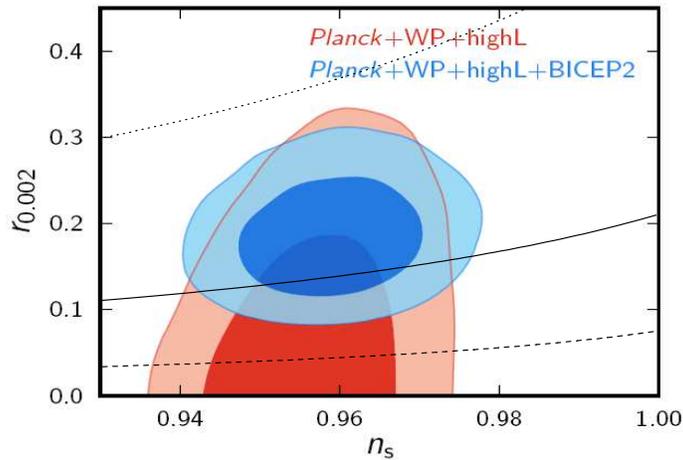}}
{\vspace{-3cm}\caption{Evolution of the tensor-scalar ratio $r$
versus the scalar spectrum index $n_s$, from the exact solution
for three different values of the parameter $\xi$. The dotted,
solid , and dashed lines are for the $\mid\xi\mid\simeq 0.10$,
$\mid\xi\mid\simeq0.08$ and $\mid\xi\mid\simeq0.06$, respectively.
 \label{fig1}}}
\end{figure}

In Fig.(\ref{fig1}) we show the evolution of  the tensor-to-scalar
ratio $r$  on the scalar spectral index $n_S$ for three different
values of $\xi$. Dotted, solid , and dashed lines are for the
$\mid\xi\mid\simeq 0.10$, $\mid\xi\mid\simeq0.08$ and
$\mid\xi\mid\simeq0.06$, respectively. Here, we note that for the
value of $\mid\xi\mid\simeq 0.10$ corresponds to $n=-1.59$ (or
equivalently $\varphi\propto t^{-1.59}$) and $p=10.04$ (or
equivalently $a\propto t^{10.04}$), see  Eqs.(\ref{p}) and
(\ref{e}). Analogously, for the value of $\mid\xi\mid\simeq0.08$
corresponds to $n=-1.7$ and $p=15.3$ and for the value of
$\mid\xi\mid\simeq0.06$ corresponds to $n=-1.8$ and $p=25.2$.

In this plot we show the two-dimensional marginalized constraints,
at 68$\%$ and 95$\%$ levels of confidence, for the
tensor-to-scalar ratio and the scalar spectral index (considered
BICEP2 experiment data \cite{B2} in connection with Planck + WP +
highL). In order to write down values for the tensor-to-scalar
ratio and the scalar spectral index,  we numerically  obtain  the
parametric plot of the consistency relation $r=r(n_S)$ considering
Eqs.(\ref{nse}) and (\ref{re}), obtained from the exact solution
for the scalar field given by Eq.(\ref{sol2}).

From this plot we find that the range for the parameter
$0.08\lesssim|\xi|\lesssim 0.10$  (or equivalently $-1.7\lesssim n
\lesssim -1.59$ and $10.04\lesssim p\lesssim 15.3$ ), which is
well supported from BICEP2 experiment. For values of $|\xi|<0.08$
the model is rejected from BICEP2, because
$r=0.2_{-0.05}^{+0.07}$, and also $r=0$ disproved at $7.0\sigma$.
Nevertheless, from Planck satellite and other CMB experiments
generated exclusively an upper limit for the tensor -to- scalar
ratio $r$, where $r<0.11$ (at 95$\%$ C.L.)\cite{Ade:2013uln}.
Recently, the Planck Collaboration has made out the data
concerning  the polarized dust emission\cite{Adam:2014bub}. From
an analysis of the  polarized thermal emissions from diffuse
Galactic dust in  different  range of frequencies,  indicates that
BICEP2  gravitational wave data could be due to the dust
contamination. Here,  an elaborated study of Planck satellite and
BICEP2 data would be demanded  for a definitive answer. In this
way, we numerically find that the parameter $|\xi|\lesssim 0.10$
is well supported by Planck satellite. Here, we note that this
constraint of $\xi$ negative is similar to found in
Ref.\cite{table}, where an effective potential $V\propto\varphi^4$
has been studied.

On the other hand,  considering  the slow roll approximation for
the  consistency relation $r=r(n_s)$ from Eqs.(\ref{nsr}) and
(\ref{rr}), we observe that the slow roll model is disproved from
observations; because the spectral index $n_S>1$, and then the
model does not work from the slow roll analysis (figure not
shown). Here, we noted that $\delta_s<0$, and then the spectral
index $n_S$ during the slow roll approximation becomes $n_S>1$.

\section{Conclusions}
In this paper we have studied the power law inflation in the
context of a non-minimally coupled scalar field. From the
equations of motion and also in the slow roll approximation we
have  found exact and slow roll solutions for our model, during
the power law expansion. In our model, we have obtained analytical
expression  for the corresponding effective potential, power
spectrum, scalar spectrum index, and tensor- to-scalar ratio
considering the exact solutions and slow roll analysis. From these
measures, we have found constraint on the parameter $|\xi|$ (or
equivalently $n$ and $p$) from BICEP2 experiment and Planck data,
where we have studied the constraint on the consistency relation
$r=r(n_s)$.

From the exact solution we have found a  constraint for the value
of the parameter $|\xi|$. In this form, from BICEP2 we have
obtained an upper bound  and a lower bound for the parameter
$|\xi|$ given by $0.08\lesssim|\xi|\lesssim 0.10$ (or equivalently
$-1.7\lesssim n \lesssim -1.59$ and $10.04\lesssim p\lesssim 15.3$
). However, we have found  that the parameter $|\xi|\lesssim0.10$
is well supported by Planck data and other CMB experiments.
Finally, we have observed during the slow-roll approximation, the
model is disproved by observations, being that the spectral index
$n_S>1$, and the model does not work from slow roll analysis.

\begin{acknowledgments}
CG and RH dedicate this article to the memory of Dr. Sergio del
Campo (R.I.P.). R.H. was supported by COMISION NACIONAL DE
CIENCIAS Y TECNOLOGIA through FONDECYT Grant N$^0$ 1130628 and
DI-PUCV N$^0$ 123.724.

\end{acknowledgments}


\end{document}